\documentclass[aps,pre,twocolumn,floatfix,superscriptaddress,showpacs,showkeys]{revtex4-1}
\usepackage{epsf,amsmath,amssymb,verbatim,color,multirow,pifont,mathrsfs,soul,amsthm, wasysym}
\usepackage{graphicx, upgreek}
\usepackage[us,12hr]{datetime}

\begin{document}

\title{Twisted states in low-dimensional hypercubic lattices}
\author{Seungjae Lee}
\affiliation{Department of Physics, Chonbuk National University, Jeonju 54896, Korea}

\author{Young Sul Cho}\email{yscho@jbnu.ac.kr}
\affiliation{Department of Physics, Chonbuk National University, Jeonju 54896, Korea}
\affiliation{Research Institute of Physics and Chemistry, Chonbuk National University, Jeonju 54896, Korea}

\author{Hyunsuk Hong}\email{hhong@jbnu.ac.kr}
\affiliation{Department of Physics, Chonbuk National University, Jeonju 54896, Korea}
\affiliation{Research Institute of Physics and Chemistry, Chonbuk National University, Jeonju 54896, Korea}

\begin{abstract}
Twisted states with non-zero winding numbers 
composed of sinusoidally coupled identical oscillators 
have been observed in a ring.
The phase of each oscillator in these states constantly shifts, following its preceding neighbor in a clockwise direction, 
and the summation of such phase shifts around the ring over $2\pi$ characterizes the winding number of each state.
In this work, we consider finite-sized $d$-dimensional hypercubic lattices, namely square ($d=2$) and cubic ($d=3$) lattices with periodic boundary conditions.
For identical oscillators, we observe new states in which the oscillators belonging to each line (plane) for $d=2$ ($d=3$)
are phase synchronized with non-zero winding numbers along the perpendicular direction.
These states can be reduced into twisted states in a ring with the same winding number if we regard
each subset of phase-synchronized oscillators as one single oscillator.
For nonidentical oscillators with heterogeneous natural frequencies, we observe similar patterns
with slightly heterogeneous phases in each line $(d=2)$ and plane $(d=3)$. We show that these states generally appear
for random configurations when the global coupling strength is larger than the critical values for the states.
\end{abstract}

\maketitle

\section{\label{sec:level1}Introduction}

Synchronization phenomena have been widely observed in a variety of real systems, such as flashing fireflies, cardiac pacemaker cells in the heart, firing neurons in the brain, coupled laser systems, electric power grids, hand clapping in concert halls, and Josephson junctions, 
among others~\cite{strogatz_sync, kurts_sync, walker_firefly, pacemaker_cell, sync_neural, sync_neural2, laser1, laser2, motter_powergrid, kuramoto_powergrid, hand_clapping, josephson}. 
The spontaneous emergence of synchronization has attracted immense interest in not only physics and biology but also many other related fields, and extensive studies have sought to understand the underlying mechanism of the phenomenon~\cite{crawford}. 
As a result, the collective properties, features, and mechanism of synchronization have been unveiled, with 
diverse patterns of synchronization and the origins of such patterns found~\cite{arenas_review}.

One representative dynamical system is the Kuramoto model, which
describes the spontaneous emergence of synchronization 
in a network of interacting oscillators~\cite{Kuramoto_1975_Nishikawa}.
In this system,
each oscillator has a natural frequency randomly assigned from probability distribution $g(\omega)$,
and the strength of coupling between each pair of connected oscillators that induces synchronization
is controlled globally via control parameter $K$.
It is known that 
order parameter $R>0$ for $K > K_c[= 2/(\pi g(0))]$ in an all-to-all coupled network structure (mean-field limit)
of infinite system size~\cite{Kuramoto_1975_Nishikawa, Kuramoto_rmp};
however, $K_c \rightarrow \infty$ as system size increases to infinity in $d$-dimensional hypercubic lattices
with Gaussian distribution $g(\omega)$ for $d \leq 4$~\cite{hhong_pre_2005}.
These results claim that low-dimensional hypercubic lattices of $d\leq4$
can be distinct from the mean-field limit.

In $(d=1)$-dimensional hypercubic lattices with periodic boundary conditions, regarded as ring structures, diverse multistable states have been reported~\cite{sync_basin_2006, sync_basin_2017, ring1, ring2, ring3, ring4, ring5}. Among these states, so-called twisted states composed of identical oscillators
were found in~\cite{sync_basin_2006, sync_basin_2017}.
In this paper, we focus on whether such twisted-state patterns
can also be observed in the low-dimensional 
hypercubic lattices of $d=2$ and $3$, which are distinct from the mean-field limit. 
In addition, we explore whether similar states are possible by using Gaussian distribution $g(\omega)$ to include heterogeneity.

The rest of this paper is organized as follows. In Sec.~\ref{sec:system},
we introduce the general system considered in this paper. In Sec.~\ref{sec:identical},
we review the twisted states in the ring and observe states of the same pattern in $d$-dimensional hypercubic lattices 
with periodic boundary conditions for $d=2$ and $3$.
In Sec.~\ref{sec:nonidentical}, we consider nonidentical oscillators with heterogeneous natural frequencies
of Gaussian distribution, where we observe similar states defined by the winding number of a cycle
and numerically check whether these states generally appear for random configurations.
We summarize the results in Sec.~\ref{sec:summary}, and provide details supporting our analysis in the Appendix.

\section{\label{sec:system}Model}
We study a Kuramoto system composed of $N$ oscillators whose phases $\phi_i \in [0, 2\pi)$ $(i=1,...,N)$ 
follow the governing equation 
\begin{equation}
\frac{d\phi_i}{dt}=\omega_i + K\sum_{j=1}^N A_{ij} \textrm{sin}(\phi_j-\phi_i)
\label{Eq:govern}
\end{equation}
for $K > 0$, where $A_{ij} = 1$ if $i$ and $j$ are connected or 0 otherwise, and
$\omega_i$ is the natural frequency of oscillator $i$.
For phase ordering, we use the order parameter given by $R=\frac{1}{N}\big|\sum_{j=1}^N e^{i\phi_j}\big|$.

In this paper, we consider $d$-dimensional hypercubic lattices for $d=2$ and $3$ of linear size $L=N^{1/d}$
with {\it{periodic boundary conditions}}.
To specify each node in the lattices, we use a Cartesian coordinate system as follows.
For a given hypercubic lattice, we choose one corner of the lattice as the origin and then define $d$ mutually 
perpendicular axes, which start from the origin and increase along its $d$ nearest neighbors.
With these axes, the location of each node can be specified using $d$ coordinates 
denoted by $(x_1,...,x_d)$. 
Here, non-negative integers for $x_1,...,x_d$ satisfy $0 \leq x_1,...,x_d\leq L-1$.

Each node at location $(x_1,...,x_d)$ is numbered by $i = \sum_{d'=1}^{d}x_{d'}L^{d'-1}+1$,
which makes it possible to analytically describe the states that we observe, as discussed later.
We note that each positive integer $i$ $(1 \leq i \leq N)$ specifies a unique location in a given lattice.
In $d$-dimensional hypercubic lattices with periodic boundary conditions, 
each node at location $(x_1,...,x_d)$ is connected with $2d$ 
nodes at locations $((x_1\pm1)~\textrm{mod}~L,...,x_d)$,...,$(x_1,...,(x_d\pm1)~\textrm{mod}~L)$.

\section{\label{sec:identical}Identical oscillators}
In this section, we consider $N$ identical oscillators with $\omega_i = \omega$ for $\forall i$.
If we use a rotating reference frame 
$\phi_i \rightarrow \phi_i + \omega t$, Eq.~(\ref{Eq:govern}) takes the form
\begin{equation}
\frac{d\phi_i}{dt}= K\sum_{j=1}^N A_{ij} \textrm{sin}(\phi_j-\phi_i).
\label{Eq:govern_identical}
\end{equation}

\begin{figure}[t!]
\includegraphics[width=1.0\linewidth]{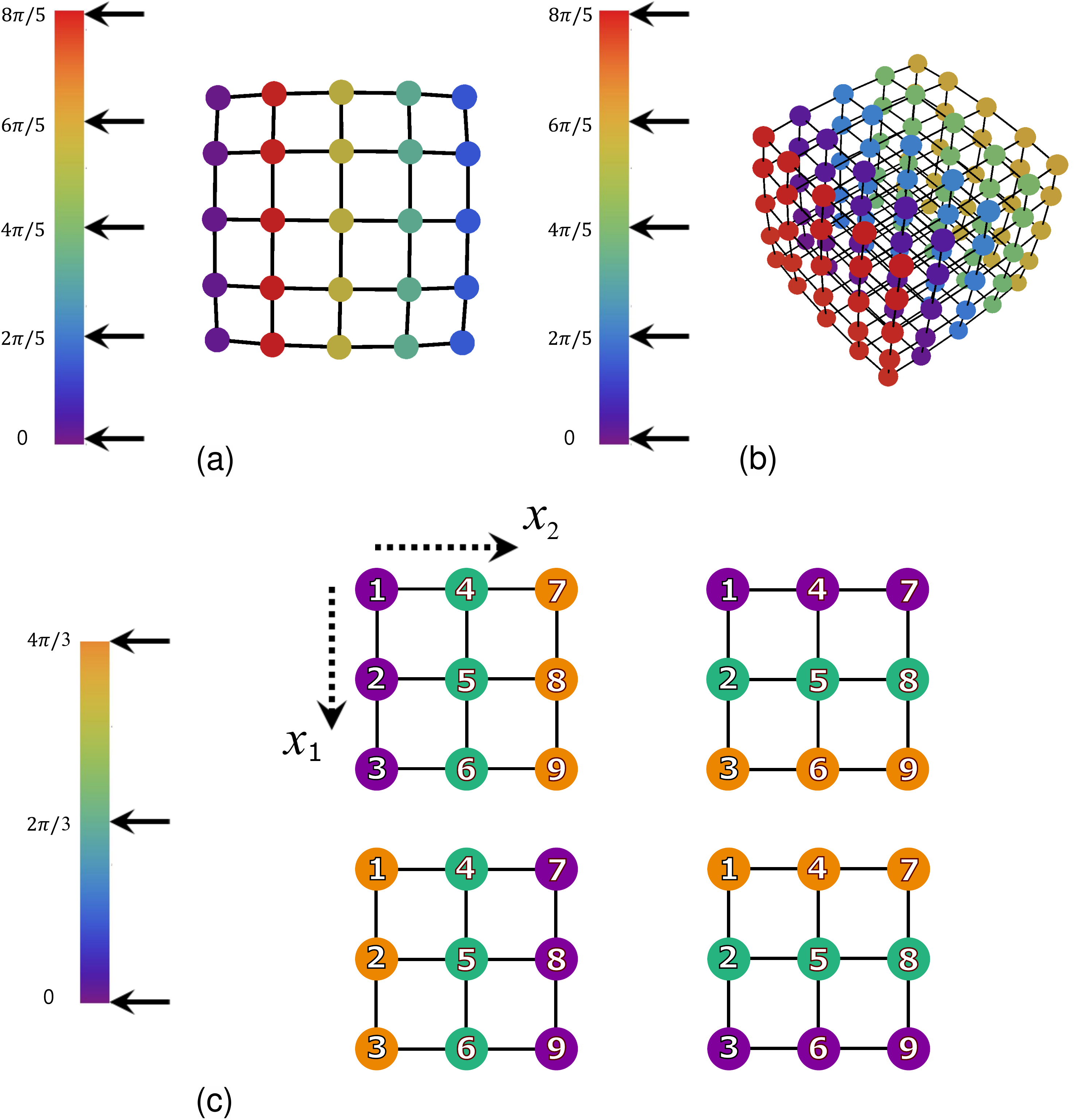}
\caption{(a, b) Schematic diagram for the ($q=1$)-twisted states in (a) two- and (b) three-dimensional hypercubic lattices for $L=5$.
The phase of each oscillator is denoted by the corresponding color of the palette on the left. 
The oscillators belonging to each column
have the same phases in (a) and the oscillators belonging to each plane have the same phases in (b).
Arrows are continually shifted upward by $2\pi/L$ starting from the bottom value $0$.
(c) Schematic diagram of $2d=4$ numbers of different patterns of $(q=1)$-twisted states for a fixed coordinate system
(dotted arrows) in a two-dimensional hypercubic lattice with $L=3$.
The phase of each oscillator is denoted by the corresponding color of the palette on the left. 
Solid arrows are continually shifted upward by $2\pi/L$ starting from the bottom value $0$.
(a-c) We note that links between pairs of nodes on opposite sides
are omitted.} 
\label{Fig:splay_schematic_L_10}
\end{figure}

\subsection{\label{subsec:twisted_ring}Twisted state on a ring}

The twisted state on a ring structure 
composed of identical oscillators was originally reported in~\cite{sync_basin_2006}.
For the ring structure, Eq.~(\ref{Eq:govern_identical}) has the form
\begin{equation}
\frac{d\phi_i}{dt}= K[\textrm{sin}(\phi_{i-1}-\phi_i)+\textrm{sin}(\phi_{i+1}-\phi_{i})]
\label{Eq:govern_identical_ring}
\end{equation} 
with $\phi_{N+1} \equiv \phi_{1}$ and $\phi_{0} \equiv \phi_{N}$. 
In a twisted state of integer winding number $q$ ($q$-twisted state), 
the phases of the oscillators are 
$\phi_i= (2\pi q i/N+C)$ mod $2\pi$ for any constant $C$.

It can be shown that this state is a {\it {fully phase-locked}} state ($\dot{\phi}_i=0$ for $\forall i$)
by substituting this form into the right-hand side of Eq.~(\ref{Eq:govern_identical_ring}).
Moreover, it has been proven that this state is linearly stable for $N > 4|q|$~\cite{sync_basin_2006, sync_basin_2017}.
Here, $q$ is called the winding number because this value refers to the number of full twists of the phase around the ring. 
We discuss the definition of winding number for an arbitrary cycle 
within the given network structure in Sec.~\ref{sec:windingnumber} \cite{ring1, ring2, ring3, ring4, ring5}.

\subsection{\label{subsec:twisted_hypercubic}Twisted states in $d$-dimensional hypercubic lattices for $d=2$ and $3$}

We observe similar states in $d$-dimensional hypercubic lattices for $d=2$ and $3$.
In these states, the oscillators in each line (plane) for $d=2$ $(d=3)$ are phase synchronized,
while the phase of each line (plane) is constantly shifted by $2\pi q/L$ 
from that of the preceding line (plane) along the perpendicular axis.

By translating and rotating the coordinate axes (renumbering the nodes following the rule in the last paragraph of Sec.~\ref{sec:system}), we can formulate these states as
\begin{equation}
\phi^*_i = \Big(\frac{2\pi q}{L}\Big\lfloor\frac{i-1}{L^{d-1}}\Big\rfloor + C\Big)~{\textrm{mod}}~2\pi
\label{Eq:twisted_generaldim}
\end{equation}
for any constant $C$ with non-negative integer $q\geq0$, 
where $\lfloor x\rfloor$ denotes the integer part of a given number $x$.
We note that the $x_d$-axis is used for the direction of non-negative winding number $q$.
We consider $0 \leq q \leq \lfloor L/2 \rfloor$
by using the restriction of the phase difference $(2\pi q/L) \in [0, \pi]$. 
We call these states {\it{$q$-twisted states}} (in $d$-dimensional hypercubic lattices)
because these states can be reduced to $q$-twisted states on a ring of size $L$ if we regard each subset of
phase-synchronized oscillators as one single oscillator with the same phase.
A schematic diagram for these states with $q=1$ is shown in Fig.~\ref{Fig:splay_schematic_L_10} (a) and (b).
We remark that all $2d$ numbers of different patterns with the same $q>0$ value for a fixed coordinate system driven by rotational and reflectional symmetry
are regarded as $q$-twisted states (Fig.~\ref{Fig:splay_schematic_L_10} (c)).

We show that the $q$-twisted state of the form Eq.~(\ref{Eq:twisted_generaldim}) 
is a fully phase-locked state of Eq.~(\ref{Eq:govern_identical}) in general $d$-dimensional
hypercubic lattices. This can be proved by inserting this form directly into the right-hand side of Eq.~(\ref{Eq:govern_identical}), 
which gives $K{\textrm{sin}}(\phi^*_{i+L^{d-1}}-\phi^*_{i})+K{\textrm{sin}}(\phi^*_{i-L^{d-1}}-\phi^*_{i})
=K{\textrm{sin}}(2\pi q/L) - K{\textrm{sin}}(2\pi q/L)=0$.

Then, we show that this state is linearly stable for $L > 4q$. For small deviations from the twisted state, 
$\phi_i = \phi^*_i + \delta \phi_i$, the rate equation for $\delta \phi_i$
up to linear order is derived as
\begin{equation}
{\delta\dot{\phi}_i}=\sum_jJ_{ij}\delta\phi_j,
\label{Eq:identical_linear_stability}
\end{equation}
where Jacobian matrix $J_{ij} \equiv \frac{\partial \dot{\phi_i}}{\partial \phi_j}\bigg|_{\phi=\phi^*}$ is given as
\begin{equation}
J_{ij} =
\begin{cases}
-K\sum_{k=1}^N A_{ik}{\textrm{cos}}(\phi^*_k-\phi^*_i) & \text{if}~~j=i, \\
KA_{ij}{\textrm{cos}}(\phi^*_j-\phi^*_i) & \text{if}~~j\neq i. \\
\end{cases}
\end{equation}
To investigate the linear stability of the twisted state, we obtain the eigenvalues of the Jacobian matrix. For $L > 4q$, 
we find that all eigenvalues are negative except for one zero, which is related to perturbation within the manifold. 
Therefore,
the twisted state is linearly stable for $L > 4q$.
For $L = 4q$, multiplicity of the zero eigenvalue is larger than one and the other eigenvalues are negative, 
which means that the twisted state is neutrally stable.
For $L < 4q$, we find that some eigenvalues are positive, and thus the twisted state 
is unstable~\cite{Marsden_jacobian, Chung_jacobian} (see Sec. A in the Appendix).

\begin{figure}[t!]
\includegraphics[width=1.0\linewidth]{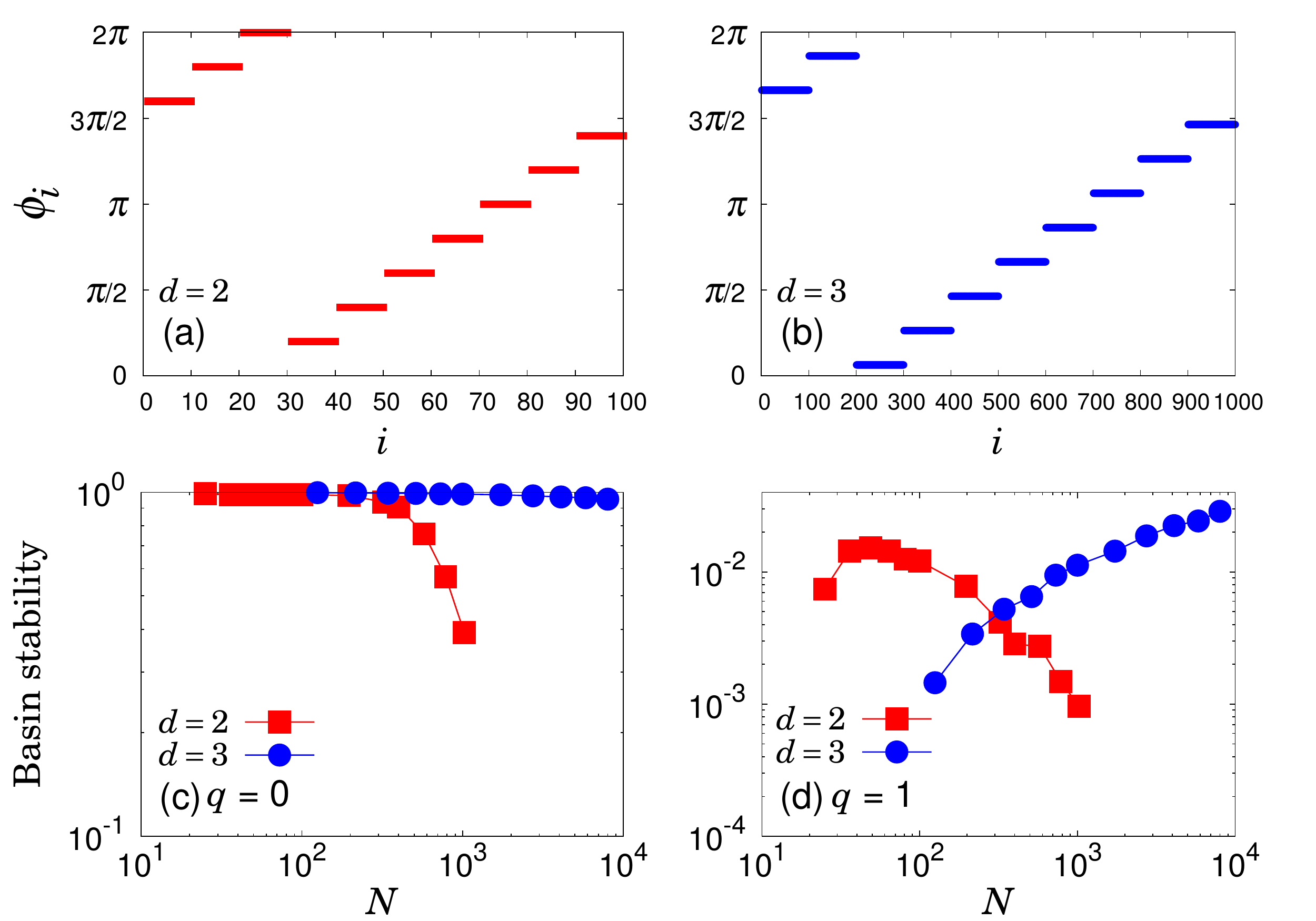}
\caption{Two configurations of numerically obtained $(q=1)$-twisted states for (a) $d=2$ and (b) $d=3$ of $L = 10$. 
Numerically measured basin stability of the twisted states of (c) $q=0$ and (d) $q=1$ for both $d=2$ and $d=3$ for various $N$.
We use $10^5$ random initial conditions with $K = 5.0$ for each $N$. We note that the result for a single $K$ value is sufficient 
because $K$ only changes the time scale in Eq.~(\ref{Eq:govern_identical}).} 
\label{Fig:twisted_phase_basin}
\end{figure}

We observe twisted states of $q=1$ for $d=2$ and $3$ by numerical simulations 
for randomly given initial phases 
$\phi_i(0) \in [0, 2\pi)$ for $\forall i$, as shown in Fig.~\ref{Fig:twisted_phase_basin}(a) and (b).
To confirm that the emergence of these twisted states is not the result of particular choices of initial phases, 
we measure the fraction of random initial conditions that induces the twisted states (referred to as the {\it {basin stability}} of the twisted states~\cite{basin_stability_kurths});
specifically, we begin with a set of random initial phases $\phi_i(0) \in [0, 2\pi)$ for $\forall i$
for each configuration. We test a large number of configurations by changing the initial phases for different configurations, and then obtain the fraction of configurations that arrives at the $q$-twisted states (basin stability of the $q$-twisted states) for each $q$ value separately.
The results for $q=0$ and $q=1$ are shown in Fig.~\ref{Fig:twisted_phase_basin}(c) and (d), respectively.
We could not observe the twisted states for $1<q<L/4$ numerically, even though the states are linearly stable.
This might be because the basin stability of the $q$-twisted states for $1<q<L/4$ are so small.

It should be noted that, for the emergence of twisted states, each pair of oscillators on opposite sides should be synchronized, and therefore the periodic boundary conditions here are important as they allow for direct coupling between the oscillators in each pair.

Now, we are curious about the spontaneous synchronization of 
the oscillators in each line ($d=2$) or in each plane ($d=3$), which seems unnatural considering the rotational symmetry of the hypercubic lattice.
We find that, in fact, such patterns originate from the translational symmetry of the hypercubic lattice~\cite{splay_symmetry}.
More precisely, we use automorphism, which is a permutation of the nodes preserving the adjacency matrix. 
In~\cite{remote_prl_2013, pecora_ncomm_2014, pecora_sciadv_2016, yscho_prl_2017},
it was reported that each set of nodes that permute to each other by an automorphism (mathematical) group can be synchronized. 
Based on such results, 
we show that synchronization of each line (plane) requires the synchronizations of all the others, 
such that every line (plane) is synchronized at the same time (see Sec. B in the Appendix).
Therefore, we expect that twisted states would be observed in other lattices that include translational symmetry,
for example hypercubic lattices for $d>3$.

\section{\label{sec:nonidentical}nonidentical oscillators}

In this section, we consider the Gaussian distribution function $g(\omega)$ with zero mean and arbitrary variances.
We choose $\omega_i$ randomly from 
a certain $g(\omega)$. 
To be specific, we choose a set $\{\omega_i\}_{1 \leq i \leq N}$ for the given $g(\omega)$ and $N$
by using the relation
\begin{equation}
\tilde{\sigma}(i) = \frac{1}{2} + N\int_{-\infty}^{\omega_i}g(\omega)d\omega,
\label{Eq:deterministic_set}
\end{equation}
where $\tilde{\sigma}$ is a random permutation of the set $\{1,2,...,N\}$. 
This method gives $\sum_{i=1}^N \omega_i=0$ exactly~\cite{hhong_pre_fss_2015}.

\subsection{\label{sec:windingnumber}Winding number of a cycle}

To find the winding number of a cycle, we consider an arbitrary cycle denoted by $c$ of length $n$ in the given network structure.
The sequence of nodes of cycle $c$ is given by $(c_0,...,c_{n-1})$. 
Then, the winding number of $c$ is given by
$q(c) = (2\pi)^{-1}\sum_{i=0}^{n-1}\Delta_{i+1, i}$,
where $\Delta_{i+1, i}=(\phi_{c_{i+1}}-\phi_{c_i})~{\textrm{mod}}~2\pi \in (-\pi, \pi]$
with $c_{n} \equiv c_0$. For dynamical systems with Eq.~(\ref{Eq:govern}), 
$q(c)$ must be an integer~\cite{ring4}.

\begin{figure}[t!]
\includegraphics[width=1.0\linewidth]{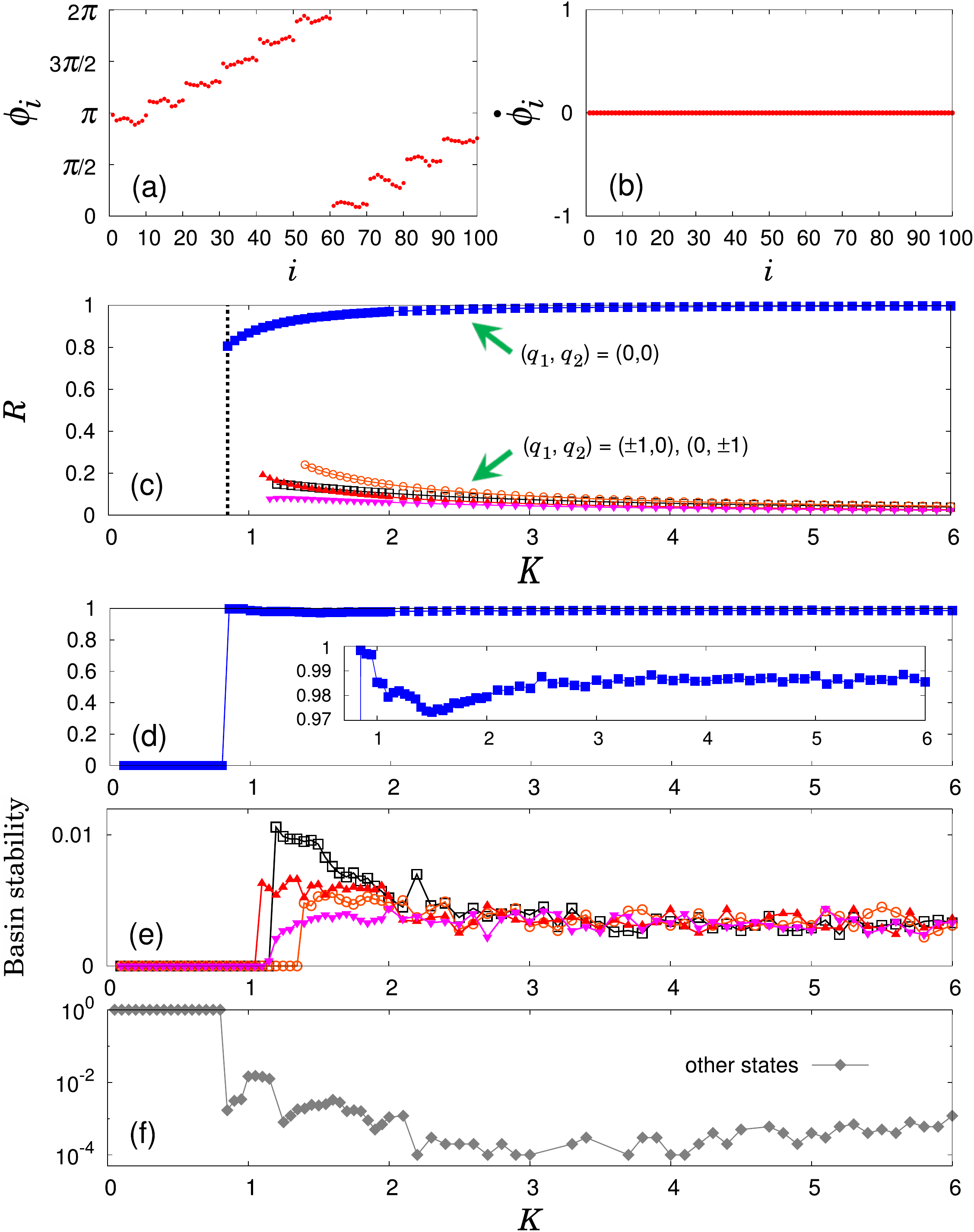}
\caption{
Data for a two-dimensional hypercubic lattice of $L=10$ for a randomly given set $\{\omega_i\}_{1\leq i\leq N}$ with unit variance, obtained by numerically integrating Eq.~(\ref{Eq:govern}) up to $t=10^3$.
(a) $\phi_i$ vs $i$ for a numerically obtained state with $(q_1,q_2)=(1,0)$. 
(b) Confirmation that the state in (a) is a fully phase-locked state. 
(c) Average value of $R$ over the numerically obtained states for pairs $(q_1,q_2)=(0,0)$ $(\blacksquare)$, 
$(1,0)$ $(\square)$, $(-1,0)$ $(\ocircle)$, $(0,1)$ $(\blacktriangle)$, and $(0,-1)$ $(\blacktriangledown)$ 
for each value of $K$. In the range of $K$ to the left of the vertical dotted line,
we cannot obtain any fully phase-locked states. 
(d-f) Numerically measured basin stabilities of states for pairs 
(d) $(q_1,q_2)=(0,0)$, (e) $(\pm1, 0)$ and $(0,\pm1)$, and (f) other states using $10^4$ random initial conditions for each value of $K$.
The inset in (d) is an enlarged plot of $R$ vs $K$ in the main panel. The symbols in (d) and (e) are the same as in (c) to denote $(q_1, q_2)$.} 
\label{Fig:Kuramoto_distorted_lattice_2dim_phase_basin_qsplit_singleseed}
\end{figure}

\begin{figure}[t!]
\includegraphics[width=1.0\linewidth]{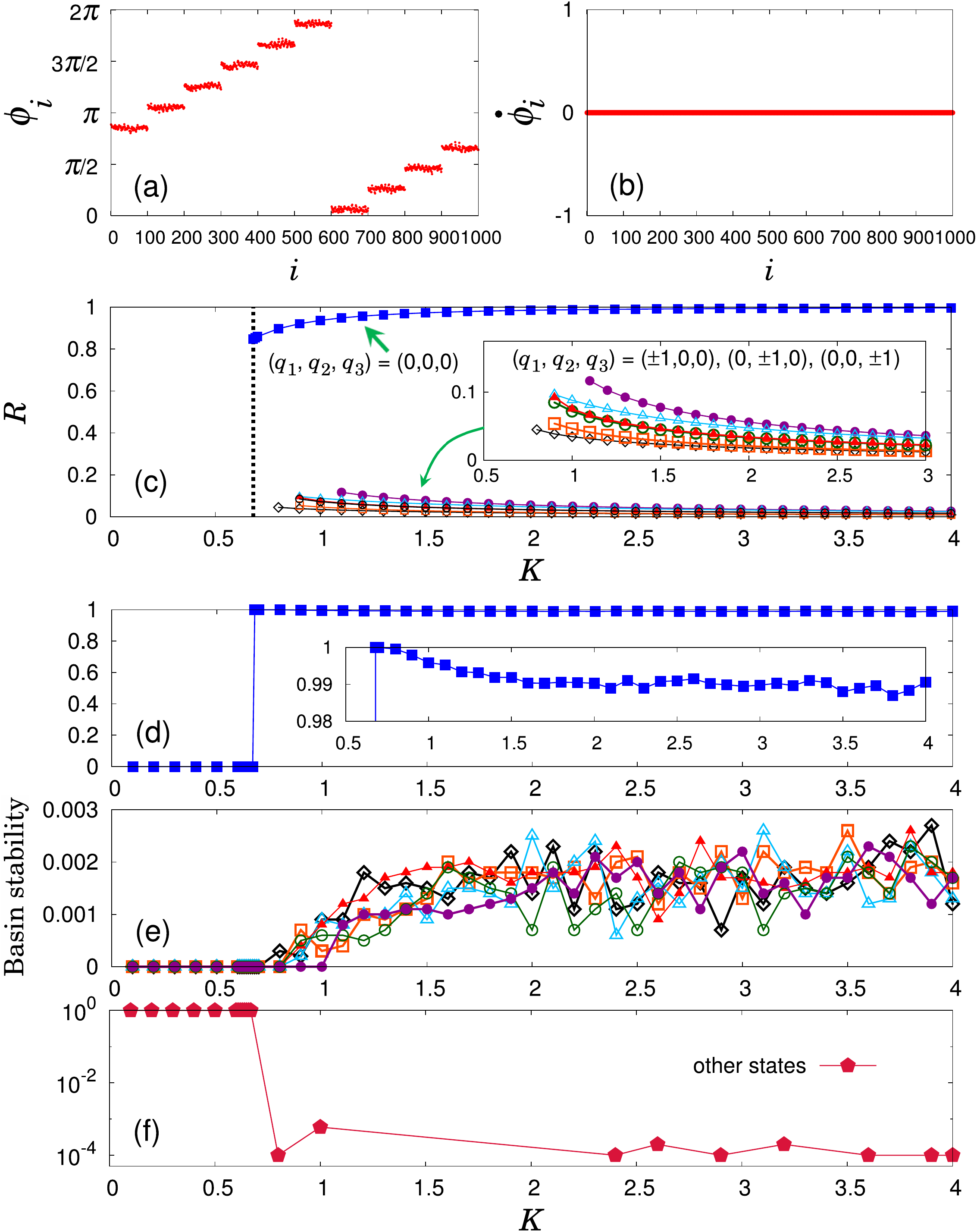}
\caption{
Data for a three-dimensional hypercubic lattice of $L=10$ for a randomly given set $\{\omega_i\}_{1 \leq i \leq N}$
with unit variance, obtained by numerically integrating Eq.~(\ref{Eq:govern}) up to $t=5 \times 10^2$.
(a) $\phi_i$ vs $i$ for a numerically obtained state with $(q_1,q_2,q_3)=(1,0,0)$. 
(b) Confirmation that the state in (a) is a fully phase-locked state. 
(c) Average value of $R$ over the numerically obtained states for $(q_1,q_2,q_3)=(0,0,0)$ $(\blacksquare)$, 
$(1,0,0)$ $(\Diamond)$, $(-1, 0, 0)$ $(\square)$, $(0,1,0)$ $(\CIRCLE)$, $(0,-1,0)$ $(\vartriangle)$, 
$(0,0,1)$ $(\blacktriangle)$, and $(0,0,-1)$ $(\Circle)$
for each value of $K$. In the range of $K$ to the left of the vertical dotted line,
we cannot obtain any fully phase-locked states. 
Inset: Enlarged plot of $R$ vs $K$ in the main panel. 
(d-f) Numerically measured basin stabilities of states for (d) $(q_1,q_2,q_3)=(0,0,0)$,
(e) $(\pm1, 0,0)$, $(0,\pm1,0)$, and $(0,0,\pm1)$, and (f) other states using $10^4$ random initial conditions for each value of $K$.
The inset in (d) is an enlarged plot of $R$ vs $K$ in the main panel. The symbols in (d) and (e) are the same as in (c) to denote $(q_1, q_2, q_3)$.
} 
\label{Fig:Kuramoto_distorted_lattice_3dim_phase_basin_qsplit_singleseed}
\end{figure}

\subsection{\label{sec:windingnumber_hypercubic}States with nonzero winding numbers along only one axis for $d=2$ and $3$}

For a $d$-dimensional hypercubic lattice with periodic boundary conditions composed of nonidentical oscillators,
we again consider the coordinate system introduced in Sec.~\ref{sec:system}.
For $d=2$ and $3$, we observe fully phased-locked states that show the same behavior as the $(q=1)$-twisted states
from the perspective of winding numbers. However, each $\phi_i$ in these states is slightly perturbed 
from Eq.~(\ref{Eq:twisted_generaldim}) in general by the heterogeneous natural frequencies.
We demonstrate these states in $d=2$ and $d=3$ lattices separately.

For $d=2$, we fix a two-dimensional coordinate system and then use notations 
$c^{(x_1,0)}$ and $c^{(0,x_2)}$ for the cycles with sequences of nodes
$(c^{(x_1,0)}_{x_2})_{x_2=0}^{L-1}$ and $(c^{(0,x_2)}_{x_1})_{x_1=0}^{L-1}$.
Here, both $c^{(x_1,0)}_{x_2}$ and $c^{(0, x_2)}_{x_1}$ denote the node at location $(x_1, x_2)$.
Therefore, each cycle $c^{(x_1,0)}$ can be regarded as a sequence of nodes reachable from $(x_1,0)$ using links along the $x_2$-axis.
Similarly, each cycle $c^{(0, x_2)}$ can be regarded as a sequence of nodes reachable from $(0, x_2)$ by paths along the $x_1$-axis.
We note that each sequence in $c^{(x_1,0)}$ and $c^{(0, x_2)}$ forms a cycle by a link connecting the two ends from the periodic boundary conditions.

We observe fully phase-locked states 
where $q(c^{(x_1,0)})=q_1$ for $\forall x_1$ and $q(c^{(0,x_2)})=q_2$ for $\forall x_2$
for five different pairs: $(q_1,q_2)=(0,0), (\pm1,0),$ and $(0,\pm1)$.
Interestingly, for $d=2$, $(q_1,q_2)=(0,0)$ is a characteristic of the $(q=0)$-twisted states,
while the other four pairs are characteristics of the $(q=1)$-twisted states.
We measure the basin stability of the states in each $(q_1, q_2)$ pair separately
for a randomly given set, $\{\omega_i\}_{1 \leq i \leq N}$, as shown in Fig.~\ref{Fig:Kuramoto_distorted_lattice_2dim_phase_basin_qsplit_singleseed}.
The numerical results support that each pair can be observed when $K$ is larger than each critical value of $K$.
We obtained similar results for $10$ different randomly given sets $\{\omega_i\}_{1\leq i \leq N}$.

For $d=3$, we fix a three-dimensional coordinate system and then use notations $c^{(x_1,x_2,0)}$, $c^{(x_1,0,x_3)}$, and $c^{(0,x_2,x_3)}$ for the cycles
with sequences of nodes $(c_{x_3}^{(x_1,x_2,0)})_{x_3=0}^{L-1}$, $(c_{x_2}^{(x_1,0,x_3)})_{x_2=0}^{L-1}$, 
and $(c_{x_1}^{(0,x_2,x_3)})_{x_1=0}^{L-1}$, respectively. Here, $c^{(x_1,x_2,0)}_{x_3}$, $c^{(x_1,0,x_3)}_{x_2}$, 
and $c^{(0,x_2,x_3)}_{x_1}$ denote the node at location $(x_1,x_2,x_3)$.
We observe fully phase-locked states 
where $q(c^{(x_1,x_2,0)})=q_1$ for $\forall {x_1, x_2}$, $q(c^{(x_1,0,x_3)})=q_2$ for $\forall x_1, x_3$, 
and $q(c^{(0,x_2,x_3)})=q_3$ for $\forall x_2, x_3$
for seven different sets: $(q_1,q_2, q_3)=(0,0,0)$, $(\pm1,0,0)$, $(0,\pm1,0)$, and $(0,0,\pm1)$.
Similar to the previous case, $(q_1,q_2,q_3)=(0,0,0)$ is a characteristic of the $(q=0)$-twisted states for $d=3$,
while the other six sets are characteristics of the $(q=1)$-twisted states.
We measure the basin stability of the states in each $(q_1,q_2,q_3)$ set 
separately for a randomly given set $\{\omega_i\}_{1\leq i \leq N}$,
with results shown in Fig.~\ref{Fig:Kuramoto_distorted_lattice_3dim_phase_basin_qsplit_singleseed}.
The numerical results again support that each set can be observed when $K$ is larger than each critical value of $K$.
We obtained similar results for $10$ different randomly given sets $\{\omega_i\}_{1\leq i \leq N}$.

In contrast to Fig.~\ref{Fig:twisted_phase_basin}(a) and (b), where $\phi_i$ vs $i$ of
each $(q=1)$-twisted state shows a clear shape, which looks like evenly spaced stairs, 
$\phi_i$ vs $i$ of the states composed of nonidentical oscillators
with $(q_1,q_2)=(1,0)$ in Fig.~\ref{Fig:Kuramoto_distorted_lattice_2dim_phase_basin_qsplit_singleseed}(a)
and $(q_1,q_2,q_3)=(1,0,0)$ in Fig.~\ref{Fig:Kuramoto_distorted_lattice_3dim_phase_basin_qsplit_singleseed}(a)
have roughness.
We ascertain that the slightly heterogeneous phases in each line 
(Fig.~\ref{Fig:Kuramoto_distorted_lattice_2dim_phase_basin_qsplit_singleseed}(a))
and plane (Fig.~\ref{Fig:Kuramoto_distorted_lattice_3dim_phase_basin_qsplit_singleseed}(a)) of the two states
originate from the heterogeneity of the natural frequencies.

\begin{figure}[t!]
\includegraphics[width=1.0\linewidth]{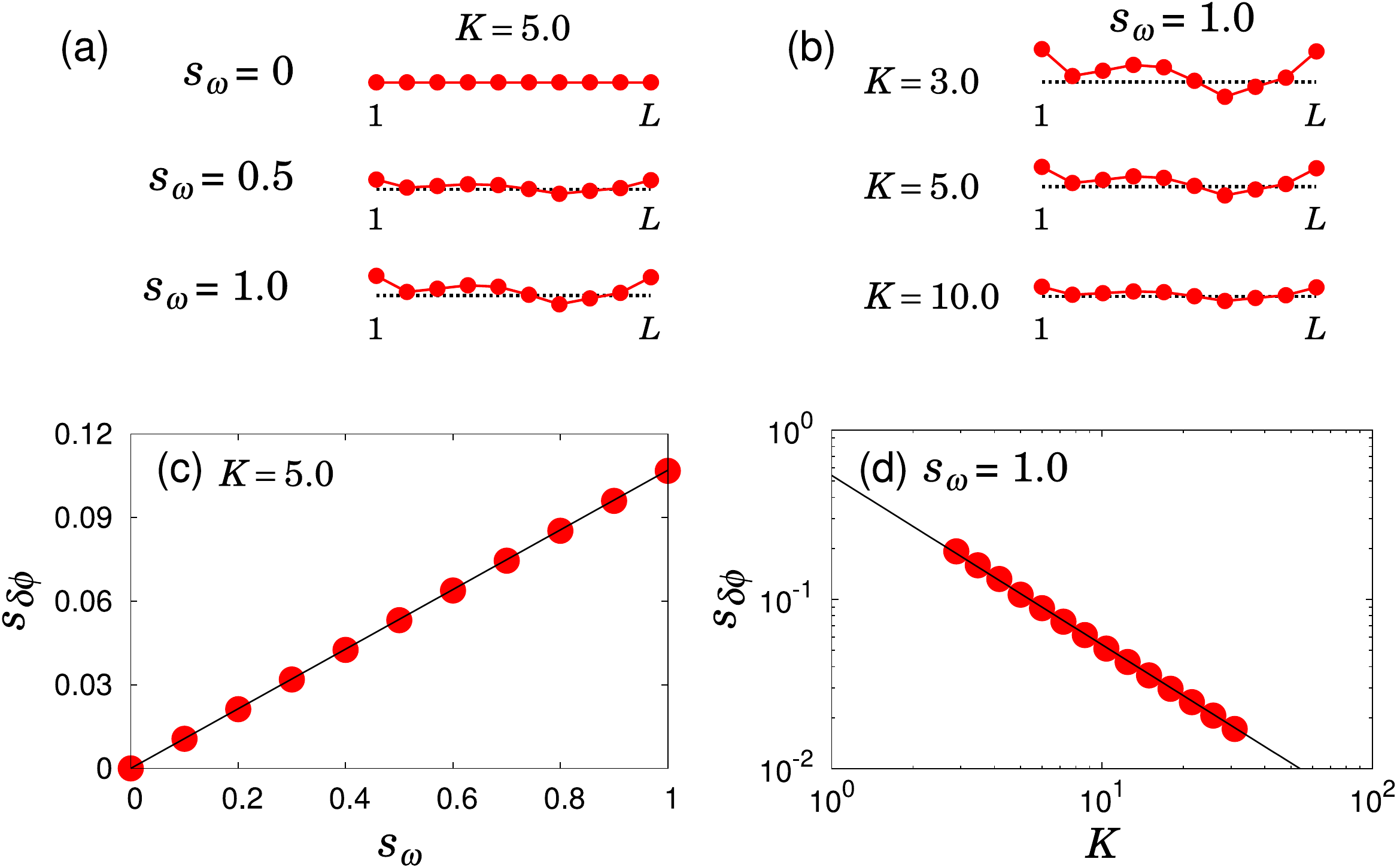}
\caption{$\phi_i$ $(\bullet)$ near $\phi^*_i$ (dotted line) of states with $(q_1,q_2)=(1,0)$
in a line in a two-dimensional lattice of $L=10$ for (a) changing $s_{\omega}$
with fixed $K = 5.0$, and for (b) changing $K$ with fixed $s_{\omega}=1.0$. 
The height of each circle denotes $\phi_i$ of corresponding $i$, with the same scale for the difference of $\phi$ to the corresponding difference of height
used for all pairs of $(s_{\omega}, K)$ in (a) and (b). 
We find that deviation from the straight form
increases as $s_{\omega}$ increases for fixed $K$ in (a), while it decreases as 
$K$ increases for fixed $s_{\omega}$ in (b). Note that the dotted line for $s_{\omega}=0$ in (a) 
is hidden because $\phi_i = \phi^*_i$ exactly. 
(c, d) Numerically obtained (c) $s_{\delta\phi}$ vs $s_{\omega}$ for fixed $K=5.0$,
and (d) $s_{\delta\phi}$ vs $K$ for fixed $s_{\omega}=1.0$.
In both (c) and (d), a two-dimensional lattice of $L=10$ is used. 
Solid lines are eye guides to show the relation $s_{\delta\phi} \propto s_{\omega}/K$.} 
\label{Fig:Kuramoto_gaussian_splaystate_stddev_K}
\end{figure}

We now consider small deviations from the $q$-twisted state given by
$\delta \phi_i = \phi_i-\phi^*_i$ with $\lvert \delta \phi_i \rvert \ll 1$ for $\forall_i$. 
Then, $\delta{\phi}_i$
in a fully phase-locked state (i.e. $\dot{\phi}_i = \delta\dot{\phi}_i=0$ for $\forall i$) follows
\begin{equation}
\omega_i=-\sum\limits_{j=1}^{N} J_{ij}\delta\phi_j.
\label{Eq:non_identical_stability}
\end{equation}
This allows us to show that $s_{\delta\phi} \propto s_{\omega}/K$ analytically, where
$s_x$ is the standard deviation of the set $\{x_i\}_{1\leq i \leq N}$.
Here, $s_{\delta\phi}=\sqrt{\sum_{i=1}^N\delta\phi_i^2/N}$ 
by using the constraint $\sum_{i=1}^N\delta\phi_i=0$, because $\sum_{i=1}^N\delta\phi_i$ can have any value 
by the singularity of the Jacobian matrix (see Sec. C in the Appendix).
We note that $s_{\omega}=\sqrt{\sum_{i=1}^N \omega_i^2/N}$ by $\sum_{i=1}^N \omega_i=0$ 
which is the result of Eq.~(\ref{Eq:deterministic_set}).
This result supports that roughness arises due to the heterogeneity of the natural frequencies.
We check $s_{\delta\phi} \propto s_{\omega}/K$ via numerical simulation
as shown in Fig.~\ref{Fig:Kuramoto_gaussian_splaystate_stddev_K}.

The result of the preceding paragraph claims that states with $(q_1,q_2)=(\pm1, 0)$, $(0,\pm1)$ for $d=2$ 
and $(q_1,q_2,q_3)=(\pm1, 0,0)$, $(0,\pm1,0)$, $(0,0,\pm1)$ for $d=3$
would become $(q=1)$-twisted states as $K \rightarrow \infty$ for a fixed $s_{\omega}$. 
Twisted states of $q > 0$ have $R=0$ exactly; consequently, $R$ of the states with these pairs of $(q_1,q_2)$ and sets of $(q_1,q_2,q_3)$
decrease to zero, as seen in Fig.~\ref{Fig:Kuramoto_distorted_lattice_2dim_phase_basin_qsplit_singleseed}(c)
and Fig.~\ref{Fig:Kuramoto_distorted_lattice_3dim_phase_basin_qsplit_singleseed}(c). 

\section{\label{sec:summary}Summary}

In summary, we studied a Kuramoto model (Eq.~(\ref{Eq:govern})) 
in two- and three-dimensional hypercubic lattices with periodic boundary conditions.
For identical oscillators in two (three) dimensions, we observed fully phase-locked states where oscillators in each line (plane) are phase synchronized, and the phase of each line (plane) is 
constantly shifted by $2\pi/L$ from that of its preceding line (plane) along the perpendicular axis.
For heterogeneous, natural frequencies given by a Gaussian distribution function, similar patterns 
with slightly heterogeneous phases in each line (plane) were observed in two (three) dimensions. 
We measured the basin stability of these states, and conclude that
such states generally appear when $K$ is larger than their critical values of $K$.

In a previous study~\cite{hhong_pre_2005}, the same system was studied using the ``annealed/adiabatic" initial condition. 
Initial phases in such a condition for each $K$ are not given randomly, but rather the phases of the steady state of the
preceding $K$ value are used successively. Therefore, the states found in this current work
may not have been observed.

\section*{Acknowledgement}
This work was supported by the NRF Grants No. 2017R1C1B1004292 (Y. S. C)
and No. 2018R1A2B6001790 (H. H).

\section*{Appendix}

\subsection{\label{appenC}Spectral properties of the Jacobian matrix to explain that the $q$-twisted state is stable only when $L>4q$}

In this section, we derive some spectral properties of the Jacobian matrix to determine 
whether the $q$-twisted state is linearly stable or not depending on $L$. 
We can obtain the Jacobian matrix for the twisted states by applying the definition 
$J_{ij}\equiv \frac{\partial\dot{\phi}_i}{\partial\phi_j}\Big|_{\phi=\phi^*}$ to Eq.~(\ref{Eq:govern_identical}), as
\begin{flalign}
&J_{ij} \equiv\frac{\partial \dot{\phi_i}}{\partial \phi_j}\bigg|_{\phi=\phi^*} = \frac{\partial}{\partial\phi_{j}}\left(K\sum\limits_{k=1}^{N}A_{ik}{\textrm{sin}}(\phi_{k}-\phi_{i})\right)\Bigg|_{\phi=\phi^*} \notag\\ &=K\sum\limits_{k=1}^{N}A_{ik}{\textrm{cos}}(\phi_k-\phi_i)(\delta_{kj}-\delta_{ij})\Bigg|_{\phi=\phi^*} \notag\\
&=KA_{ij}{\textrm{cos}}(\phi^{*}_j-\phi^{*}_i)-\delta_{ij}K\sum\limits_{k=1}^{N}A_{ik}{\textrm{cos}}(\phi^{*}_k-\phi^{*}_i)\notag\\
&=\begin{cases}
-K\sum_{k=1}^{N}A_{ik}{\textrm{cos}}(\phi^{*}_k-\phi^{*}_i) & \text{if}~~j=i, \\
KA_{ij}{\textrm{cos}}(\phi^{*}_j-\phi^{*}_i) & \text{if}~~j\neq i.
\end{cases}
\label{Eq:Jacobian_matrix1}
\end{flalign}
From Eq.~(\ref{Eq:Jacobian_matrix1}), we can confirm that $J_{ij}=J_{ji}$, or in other words, $\bold{J}$ is a real symmetric matrix. 
Therefore $\bold{J}$, is orthogonally diagonalizable.
In order to investigate the linear stability of the $q$-twisted state, 
we need to obtain some information on the eigenvalues of the Jacobian matrix.

\subsubsection{\label{appendC1}Non-positive eigenvalues in the Jacobian matrix for $L \geq 4q$}
We show that the Jacobian matrix for $L \geq 4q$ is a negative semi-definite matrix whose eigenvalues are non-positive. 
The incidence matrix $\bold{M}$ is a $Nd \times N$ matrix whose elements are 
$M_{ei}=\sqrt{KA_{ij}{\textrm{cos}}(\phi^*_j-\phi^*_i)}$ and $M_{ej}=-\sqrt{KA_{ij}{\textrm{cos}}(\phi^*_j-\phi^*_i)}$ 
if $e$ connects two nodes $i$ and $j$ but $0$ otherwise. 
From the definition of the incidence matrix, we can derive 
\begin{flalign}
[\bold{M}^{\top}\bold{M}]_{ii} = \sum\limits_{e=1}^{Nd} (M_{ei})^2 =
\sum\limits_{j=1}^{N} KA_{ij}{\textrm{cos}}(\phi^*_j-\phi^*_i)=-J_{ii}, \notag
\end{flalign}
\begin{flalign}
[\bold{M}^{\top}\bold{M}]_{ij} = \sum\limits_{e=1}^{Nd} M_{ei}M_{ej} =
-KA_{ij}{\textrm{cos}}(\phi^*_j-\phi^*_i)=-J_{ij}.
\label{Eq:incidence_jacobian}
\end{flalign}

From Eq.~(\ref{Eq:incidence_jacobian}), ${\bold J}=-{\bold M}^{\top}{\bold M}$. 
If we use ${\bold v}_i$ and $\lambda_i$ for the $i$-th eigenvector and 
eigenvalue of $\bold{J}$, $\bold{J}{\bold v}_i=\lambda_i{\bold v}_i$ for $i=1,...,N$.
Here, ${\bold v}_i$ are real vectors and $\lambda_i$ are real values because $\bold{J}$ is orthogonally diagonalizable. 
Then, $\lambda_i={\bold v}^{\top}_i{\bold J}{\bold v}_i=-{\bold v}^{\top}_i{\bold M}^{\top}{\bold M}{\bold v}_i
=-|{\bold M}{\bold v}_i|^2 \leq0$ for $\forall_i$ when $L \geq 4q$.
We remark that this is not applicable for $L < 4q$ (i.e. $L/4 < q \leq \lfloor L/2 \rfloor$) because some elements of $\bold{M}$, $\sqrt{K{\textrm{cos}}(2\pi q/L)}$ 
are imaginary numbers. Therefore, $\bold{J}$ for $L \geq 4q$ is a negative semi-definite matrix and its eigenvalues are non-positive.

\subsubsection{\label{appenC2}Zero eigenvalue of multiplicity one for $L > 4q$ and multiplicity $L$ for $L = 4q$ in the Jacobian matrix}

In this subsection, we consider negative semi-definite matrix $\bold{J}$ for $L \geq 4q$.
$\bold{J}$ has at least one eigenvalue of $0$ related to stability for the perturbation of constant phase shifts for all oscillators. 
Therefore, the twisted state is stable if the multiplicity of eigenvalue $0$ is one, but the twisted state is neutrally stable if the multiplicity of eigenvalue $0$ is larger than one.
We show that the twisted state is stable for $L > 4q$ and neutrally stable for $L=4q$ by investigating the multiplicity in each case.

For the analysis, we decompose the hypercubic lattice into two sub-networks whose adjacency matrices are denoted by ${\bold A}^{(1)}$ and ${\bold A}^{(2)}$. The first sub-network consists of $2(d-1)N$ number of links parallel to the $x_{d'}$-axis for $1\leq d'\leq d-1$, 
and the second sub-network consists of the other $2N$ number of links parallel to the $x_d$-axis.
Then, ${\bold J}$ is decomposed into two terms as
\begin{equation}
{\bold J} = -K{\bold L}^{(1)} - K{\textrm{cos}\Big(\frac{2\pi q}{L}\Big)}{\bold L}^{(2)},
\end{equation}
where ${\bold L}^{(1)}$ and ${\bold L}^{(2)}$ are the Laplacian matrices of ${\bold A}^{(1)}$ and ${\bold A}^{(2)}$, respectively.

When $L=4q$, the Jacobian matrix is reduced to ${\bold J}=-K{\bold L}^{(1)}$. 
It is known that the multiplicity of the eigenvalue $0$ in a Laplacian matrix is equal to the number of connected components 
in the given network. The number of connected components in ${\bold A}^{(1)}$ is $L$, 
which clarifies that the multiplicity of eigenvalue $0$ is $L=4q$. Therefore, the twisted state for $L=4q$ is neutrally stable.

On the other hand, when $L>4q$ we show that the multiplicity of eigenvalue $0$ is one by way of contradiction. 
We assume that eigenvalue $0$ has two linearly independent eigenvectors ${\bold v}_1$ and ${\bold v}_2$.
Then,
\begin{flalign}
0&= {\bold v}^{\top}_1{\bold J}{\bold v}_1 = -{\bold v}^{\top}_1 \left(K{\bold L}^{(1)}+K\cos \left( \frac{2\pi q}{L} \right) {\bold L}^{(2)} \right) {\bold v}_1\notag\\
&=-K{\bold v}^{\top}_1 {\bold L}^{(1)} {\bold v}_1 -K\cos \left( \frac{2\pi q}{L} \right){\bold v}^{\top}_1{\bold L}^{(2)} {\bold v}_1\notag\\
&=-\frac{K}{2}\sum_{i,j=1}^NA^{(1)}_{ij}\left(v_{1i}-v_{1j} \right)^2\notag\\
&-\frac{K}{2}\cos \left( \frac{2\pi q}{L} \right)\sum_{i, j = 1}^NA^{(2)}_{ij}\left(v_{1i}-v_{1j} \right)^2 = 0.
\end{flalign}
Since each term in the summations is positive, i.e. ${\bold v}_1$ is a real vector because $\bold{J}$ is orthogonally diagonalizable,
each term should be zero. Thus, $v_{1i} = v_{1j}$ for $\forall_{i,j}$ because all the oscillators belong to one connected component
for $\bold{A} = \bold{A}^{(1)} + \bold{A}^{(2)}$. In the same way, we can obtain $v_{2i} = v_{2j}$ for $\forall_{i,j}$, 
which tells us that ${\bold v}_2$ is a scalar multiplication of ${\bold v}_1$. 
This violates the assumption that ${\bold v}_1$ and ${\bold v}_2$ are linearly independent, thereby proving that
multiplicity of eigenvalue $0$ is one for $L > 4q$. Therefore, the twisted state for $L>4q$ is stable.

\subsubsection{\label{appenC3} Positive eigenvalues in the Jacobian matrix for $L < 4q$}

In this subsection, we show that at least $(L-1)$ eigenvalues of $\bold{J}$ for $L < 4q$ are positive.
We consider $N \times 1$ column vectors $\bold{v}^{(\ell)}$ $(\ell=0,...,L-1)$ where 
$v^{(\ell)}_i=1/\sqrt{L^{d-1}}$ if $\lfloor(i-1)/L^{d-1}\rfloor = \ell$ and $0$ otherwise.
$\bold{v}^{(\ell)}$ is parallel to the direction of the constant phase shifts of the nodes in the $\ell$-th synchronized subset.
Then, ${\bold J}\bold{v}^{(\ell)}=-K{\textrm{cos}}(2\pi q/L){\bold L}^{(2)}\bold{v}^{(\ell)}
=K{\textrm{cos}}(2\pi q/L)(\bold{v}^{(\ell+1)}-2\bold{v}^{(\ell)}+\bold{v}^{(\ell-1)})$
with $\bold{v}^{(L)} \equiv \bold{v}^{(0)}$ and $\bold{v}^{(-1)} \equiv \bold{v}^{(L-1)}$.

An $L \times L$ matrix $\tilde{\bold{J}}^{(2)}$ given by
\begin{flalign}
\tilde{J}^{(2)}_{mn} = 
\begin{cases}
-2K{\textrm{cos}}\big(\frac{2\pi q}{L}\big) & \text{if}~~n=m, \\
K{\textrm{cos}}\big(\frac{2\pi q}{L}\big) & \text{if}~~n=1+(m\pm 1)~{\textrm{mod}}~L\\
0~~~ & \text{otherwise}
\label{Eq:Jacobian_matrix_ring}
\end{cases}
\end{flalign}
is orthogonally diagonalizable for an orthogonal matrix $\bold{U}$, such as $\bold{U}^{\top}\tilde{\bold{J}}^{(2)}\bold{U}=\bold{\Lambda}$
where $\varLambda_{mn}=\lambda_{n}\delta_{mn}$.
Without a loss of generality, we set the first eigenvalue to $\lambda_1=0$ by fixing the first column of $\bold{U}$ as $U_{m1}=\frac{1}{\sqrt{L}}$.
Here, $\tilde{\bold{J}}^{(2)}=-K\textrm{cos}(2\pi q/L)\tilde{\bold{L}}^{(2)}$
for the Laplacian matrix of a ring of length $L$ denoted by $\tilde{\bold{L}}^{(2)}$.
By the property of the Laplacian matrix, $\tilde{\bold{L}}^{(2)}$ has one zero-eigenvalue 
and all the other eigenvalues are positive. 
Therefore, $\lambda_n > 0$ ($1<n\leq L$) for $L < 4q$. 
Finally, $\bold{J}\big(\sum_{m=1}^{L}U_{mn}{\bold v}^{(m-1)}\big)=\lambda_n\big(\sum_{m=1}^{L}U_{mn}{\bold v}^{(m-1)}\big)$,
which means that $\sum_{m=1}^{L}U_{mn}{\bold v}^{(m-1)}$ is an eigenvector of $\bold{J}$ with the eigenvalue $\lambda_n$.
Therefore, at least $(L-1)$ positive eigenvalues (i.e., $\lambda_n$ for $1 < n \leq L$) of $\bold{J}$ exist.

\subsection{\label{appenB}Symmetry-inducing synchronization of subsets in the twisted states}

In this section, we analyze the relation between the automorphism and synchronization of the subset of oscillators 
in the twisted states for both two and three dimensions in more detail.

For $d=2$, we consider a permutation of nodes $\sigma$ given by
\begin{equation}
\sigma(i) = 
\begin{cases}
i-L+1 & \text{if}~~i~\text{mod}~L=0, \\
i + 1 & \textrm{otherwise}.
\end{cases}
\end{equation}
Under this permutation, the adjacency matrix is preserved (i.e., $A_{ij}=A_{\sigma(i)\sigma(j)}$), which means that
$\sigma$ is an automorphism for $d=2$. Then we can construct the group $G=\big<\sigma\big>$ generated by $\sigma$.
The orbit of $i$ operated by $G$, denoted by $\varphi(G,i)$, is defined as $\varphi(G, i)=\{\tilde{\sigma}(i) | \tilde{\sigma} \in G\}$.
By the property of the group, it is guaranteed that $\varphi(G, i)=\varphi(G, j)$ for all $j \in \varphi(G, i)$.
Here, $\varphi(G, i)$ is the subset of oscillators belonging to the same line with $i$, and 
all oscillators in the lattice are partitioned into a unique set of orbits
$\{\varphi(G, 1), \varphi(G, L+1), ..., \varphi(G, L(L-1)+1)\}=\{\varphi_1,...,\varphi_{\ell},...,\varphi_L\}$ by $G$.
Here, reduced notation $\varphi_{\ell}$ denotes the subset of oscillators in the $\ell$-th orbit, 
which is the same as the subset of oscillators in the $\ell$-th line.

For $d=3$, we consider two automorphisms $\sigma_1$ and $\sigma_2$ given by
\begin{flalign}
\sigma_1(i) =
\begin{cases}
i-L+1 & \text{if}~~i~\text{mod}~L=0,\\
i+1 & \text{otherwise,}
\end{cases} \notag
\end{flalign}
and
\begin{flalign}
\sigma_2(i) =
\begin{cases}
i-(L-1)L & \text{if}~~\{1+(i-i~{\textrm{mod}}~L)/L\}~\textrm{mod}~L=0,\\
i+L & \text{otherwise.}
\end{cases} \notag
\end{flalign}
For the group $G=\big<\sigma_1,\sigma_2\big>$ generated by $\sigma_1$ and $\sigma_2$, all oscillators 
in the lattice are partitioned into a unique set of orbits
$\{\varphi(G, 1), \varphi(G, L^2+1), ..., \varphi(G, L^2(L-1)+1)\}=\{\varphi_1,...,\varphi_{\ell},...,\varphi_L\}$ by $G$,
where $\varphi_{\ell}$ is the same as the subset of the oscillators in the $\ell$-th plane.
In both dimensional cases, we consider general states where $\phi_i=s_{\ell}$ for $\forall$$i \in \varphi_{\ell}$ $(\ell=1,...,L)$. 
Then, an arbitrary $i \in \varphi_{\ell}$ follows the equation
\begin{equation}
\dot{s}_{\ell} = K{\textrm {sin}}(s_{\ell+1}-s_{\ell}) + K{\textrm {sin}}(s_{\ell-1}-s_{\ell}).
\end{equation}
This means that each subset of oscillators can be synchronized when all the other subsets are synchronized.
In this way, symmetry can explain the simultaneous synchronizations of all subsets.

\subsection{\label{appenD}Derivation of $s_{\delta \phi} \propto s_{\omega}/K$}

In this section, we derive $s_{\delta \phi} \propto s_{\omega}/K$
starting from ${\boldsymbol{\upomega}}=-{\bold J}\delta\boldsymbol{\upphi}$ (Eq.~(\ref{Eq:non_identical_stability})),
where Eq.~(\ref{Eq:non_identical_stability}) is represented by using the vectors $\boldsymbol{\upomega}=(\omega_1,...,\omega_N)^{\top}$ 
and $\delta\boldsymbol{\upphi}=(\delta\phi_1,...,\delta\phi_N)^{\top}$. 
We consider $\bold{J}$ for $L > 4q$.
$\bold{J}$ is orthogonally diagonalizable for an orthogonal matrix $\bold{U}$, such as 
$\bold{U}^{\top}\bold{J}\bold{U}=\bold{\Lambda}$,
where $\varLambda_{ij}=\lambda_{i}\delta_{ij}$ and $\lambda_i$ is the $i$-th eigenvalue of $\bold{J}$.
We showed that only one zero-eigenvalue exists for $\bold{J}$ with $L > 4q$ in Sec.~B. 
Without a loss of generality, we set the first eigenvalue to $\lambda_1=0$ and $\lambda_i\neq0$ for $i > 1$. 
Under these conditions, the first column of $\bold{U}$ is fixed as $U_{i1} = \frac{1}{\sqrt{N}}$.

From $\boldsymbol{\upomega}=-{\bold J}\delta\boldsymbol{\upphi}$, we can derive 
${\bold U}^{\top}\boldsymbol{\upomega}=-\bold{\Lambda}{\bold U}^{\top}\delta\boldsymbol{\upphi}$
by ${\bold U}^{\top}={\bold U}^{-1}$, which comes from the orthogonality of $\bold U$. This equation can be written component-wise as
\begin{flalign}
\left[{\bold U}^{\top}\boldsymbol{\upomega}\right]_i =-\lambda_i\big[{\bold U}^{\top}\delta\boldsymbol{\upphi}\big]_i.
\label{Eq:S7}
\end{flalign}
For $i=1$, Eq.~(\ref{Eq:S7}) can be written as 
$\sum_j\omega_j = -\lambda_1 \sum_j \delta\phi_j$. 
Therefore, $\sum_j\delta\phi_j$ can have any value because $\sum_j\omega_j=0$ and $\lambda_1=0$.
This property might be given by the singularity of $\bold{J}$.
We use constraint $\sum_j\delta\phi_j=0$ for later use. In fact, we checked whether 
$\boldsymbol{\upomega}=-{\bold J}\delta\boldsymbol{\upphi}$ is solvable by using the pseudo-inverse property of $\bold{J}$, with the result giving a
comparable solution with the numerical data under this constraint.

For $i > 1$ with $\lambda_i\neq0$, Eq.~(\ref{Eq:S7}) can be transformed into
$-\left[{\bold U}^{\top}\boldsymbol{\upomega}\right]_i/\lambda_i =\big[{\bold U}^{\top}\delta\boldsymbol{\upphi}\big]_i$.
By using $\delta\boldsymbol{\upphi}^{\top}\delta\boldsymbol{\upphi}=({\bold U}^{\top}\delta\boldsymbol{\upphi})^{\top}({\bold U}^{\top}\delta\boldsymbol{\upphi})$,
we can derive the relation between $s_{\delta\phi}$ and $\boldsymbol{\upomega}$ as
\begin{flalign}
s_{\delta\phi}=&
\sqrt{\frac{1}{N}\sum\limits_{i=1}^{N}\delta\phi_i^2}\notag \\
&=\sqrt{\frac{1}{N}\sum\limits_{i=1}^{N}[{\bold U}^{\top}\delta\boldsymbol{\upphi}]^2_i} \notag \\
&=\frac{1}{\sqrt{N}}\sqrt{\sum_{i=2}^{N}\frac{1}{\lambda_i^2}[{\bold U}^{\top}\boldsymbol{\upomega}]_i^2
+\frac{1}{N}\Big(\sum_{i=1}^{N}\delta\phi_i\Big)^2} \notag\\
&=\frac{1}{\sqrt{N}}\sqrt{\sum_{i=2}^{N}\frac{1}{\lambda_i^2}[{\bold U}^{\top}\boldsymbol{\upomega}]_i^2}.
\label{Eq:S8}
\end{flalign}
For the last step, we use the constraint $\sum_{i=1}^{N}\delta\phi_i=0$.

For an arbitrary random permutation $\tilde{\sigma}$ in Sec.~\ref{sec:nonidentical},
we change $s_{\omega}$ for a fixed $\tilde{\sigma}$. 
Then, $\boldsymbol{\upomega}$ is a function
of $\tilde{\sigma}$ and $s_{\omega}$ given by $\boldsymbol{\upomega}(\tilde{\sigma}, 
s_{\omega})=s_{\omega}\boldsymbol{\upomega}(\tilde{\sigma}, 1)$.
If we average Eq.~(\ref{Eq:S8}) over all permutations $\tilde{\sigma}$, the right hand side is represented as
\begin{flalign}
s_{\delta\phi}=&
\frac{1}{\sqrt{N}}\left\langle
\sqrt{\sum_{j=2}^{N}\frac{1}{\lambda_j^2}[{\bold U}^{\top}\boldsymbol{\upomega}(\tilde{\sigma}, s_{\omega})]_j^2}
\right\rangle_{\tilde{\sigma}} \notag\\
=&\frac{1}{\sqrt{N}}\left\langle
\sqrt{\sum_{j=2}^{N}\frac{1}{\lambda_j^2}[{\bold U}^{\top}{s_{\omega}}\boldsymbol{\upomega}(\tilde{\sigma}, 1)]_j^2}
\right\rangle_{\tilde{\sigma}} \notag \\
=&\frac{s_{\omega}}{\sqrt{N}}\left\langle
\sqrt{\sum_{j=2}^{N}\frac{1}{\lambda_j^2}[{\bold U}^{\top}\boldsymbol{\upomega}(\tilde{\sigma}, 1)]_j^2}
\right\rangle_{\tilde{\sigma}} \notag \\
\propto&~\frac{s_{\omega}}{K}
\label{Eq:S9}
\end{flalign}
for fixed $N$. Ultimately, we use $\lambda_i = K\overline{\lambda}_i$ for $\forall i$ where $\overline{\lambda}_i$
is the $i$-th eigenvalue of $\bold{J}$ for $K=1$. 
Accordingly, we derived the relation $s_{\delta\phi} \propto s_{\omega}/K$,
which we numerically checked as shown in Fig.~\ref{Fig:Kuramoto_gaussian_splaystate_stddev_K}.


\vskip 1cm 
\begin{thebibliography}{99}
\bibitem{strogatz_sync} S. H. Strogatz, {\it {Sync}} (Hyperion, New York, 2003).
\bibitem{kurts_sync} A. Pikovsky, M. Rosenblum, and J. Kurths, {\it {Synchronization: A Universal Concept in Nonlinear Sciences}}
(Cambridge University Press, Cambridge, England, 2001).
\bibitem{walker_firefly} T. J. Walker, Science {\bf 166}, 891 (1969).
\bibitem{pacemaker_cell} C. Liu, D. R. Weaver, S. H. Strogatz, and S. M. Reppert, Cell {\bf 91}, 855 (1997).
\bibitem{sync_neural} F. Varela, J.-P. Lachaux, E. Rodriguez, and J. Martinerie, Nat. Rev. Neurosci. {\bf 2}, 229 (2001).
\bibitem{sync_neural2} A. K. Engel, P. Fries, and W. Singer, Nat. Rev. Neurosci. {\bf 2}, 704 (2001).
\bibitem{laser1} Z. Jiang and M. McCall, J. Opt. Soc. Am. B {\bf 10}, 155 (1993).
\bibitem{laser2} S. Yu. Kourtchatov, V. V. Likhanskii, A. P. Napartovich, F. T. Arecchi, and A. Lapucci, Phys. Rev. A {\bf 52}, 4089 (1995).
\bibitem{motter_powergrid} A. E. Motter, S. A. Myers, M. Anghel, and T. Nishikawa, Nat. Phys. {\bf 9}, 191 (2013).
\bibitem{kuramoto_powergrid} G. Filatrella, A. H. Nielsen, and N. F. Pedersen, Eur. Phys. J. B {\bf 61}, 485 (2008).
\bibitem{hand_clapping} Z. N\'eda, E. Ravasz, Y. Brechet, T. Vicsek, and A.-L. Barab\'asi, Nature {\bf 403}, 849 (2000).
\bibitem{josephson} K. Wiesenfeld, P. Colet, and S. H. Strogatz, Phys. Rev. Lett. {\bf 76}, 404 (1996).
\bibitem{crawford} J. D. Crawford, J. Stat. Phys. {\bf 74}, 1047 (1994); J. D. Crawford, Phys. Rev. Lett. {\bf 74}, 4341 (1995);
J. D. Crawford and K. T. R. Davies, Physica D {\bf 125}, 1 (1999).
\bibitem{arenas_review} A. Arenas, A. D.-Guilera, J. Kurths, Y. Moreno, and C. Zhou, Phys. Rep. {\bf 469}, 93 (2008).
\bibitem{Kuramoto_1975_Nishikawa} Y. Kuramoto, in {\it{Proceedings of the International Symposium on Mathematical Problems in Theoretical Physics,}}
edited by H. Araki (Springer-Verlag, New York, 1975); Y. Kuramoto and I. Nishikawa, J. Stat. Phys. {\bf 49}, 569 (1987).
\bibitem{Kuramoto_rmp} J. A. Acebr\'{o}n, L. L. Bonilla, C. J. P\'{e}rez Vicente, and F. Ritort, Rev. Mod. Phys. {\bf 77}, 137 (2005).
\bibitem{hhong_pre_2005} H. Hong, H. Park, and M. Y. Choi, Phys. Rev. E {\bf 72}, 036217 (2005).
\bibitem{sync_basin_2006} D. A. Wiley, S. H. Strogatz, and M. Girvan, Chaos {\bf 16}, 015103 (2006).
\bibitem{sync_basin_2017} R. Delabays, M. Tyloo, and P. Jacquod, Chaos {\bf 27}, 103109 (2017).
\bibitem{ring1} J. A. Rogge and D. Aeyels, J. Phys. A {\bf 37}, 11135 (2004).
\bibitem{ring2} R. Delabays, T. Coletta, and P. Jacquod, J. Math. Phys. {\bf 57}, 032701 (2016).
\bibitem{ring3} J. Ochab and P. F. G\'ora, Acta Phys. Pol. B Proc. Suppl. {\bf 3}, 453 (2010).
\bibitem{ring4} R. Delabays, T. Coletta, and P. Jacquod, J. Math. Phys. {\bf 58}, 032703 (2017).
\bibitem{ring5} D. Manik, M. Timme, and D. Witthaut, Chaos {\bf 27}, 083123 (2017).
\bibitem{Marsden_jacobian} A. Marsden, {\it{Eigenvalues of the Laplacian and Their Relationship to the Connectedness of a Graph}} (University of Chicago, REU, 2013). 
\bibitem{Chung_jacobian} F. R. K. Chung, {\it{Spectral Graph Theory}} (AMS, Providence, RI, 1997).
\bibitem{basin_stability_kurths} P. J. Menck, J. Heitzig, N. Marwan, and J. Kurths, Nat. Phys. {\bf 9}, 89 (2013).
\bibitem{splay_symmetry} R. E. Mirollo, SIAM J. Math. Anal. {\bf 25}, 1176 (1994).
\bibitem{remote_prl_2013} V. Nicosia, M. Valencia, M. Chavez, A. D\'{\i}az-Guilera, and V. Latora, Phys. Rev. Lett. {\bf 110}, 174102 (2013).
\bibitem{pecora_ncomm_2014} L. M. Pecora, F. Sorrentino, A. M. Hagerstrom, T. E. Murphy, and R. Roy, Nat. Commun. {\bf 5}, 4079 (2014).
\bibitem{pecora_sciadv_2016} F. Sorrentino, L. M. Pecora, A. M. Hagerstrom, T. E. Murphy, and R. Roy, Sci. Adv. {\bf 2}, e1501737 (2016).
\bibitem{yscho_prl_2017} Y. S. Cho, T. Nishikawa, and A. E. Motter, Phys. Rev. Lett. {\bf 119}, 084101 (2017).
\bibitem{hhong_pre_fss_2015} H. Hong, H. Chat\'e, L.-H. Tang, and H. Park, Phys. Rev. E {\bf 92}, 022122 (2015).




\end{thebibliography}
\end{document}